\documentclass{optica-article}

\journal{opticajournal} 

\articletype{Research Article}

\usepackage{lineno}
\usepackage{graphicx}
\usepackage{subcaption}
\usepackage{soul}
\usepackage{float}
\usepackage{enumitem}

\begin{document}

\title{Baudrate- and Reach-Flexible All-Optical Equalization with a Co-Packaged Photonic Reservoir and Receiver}

\author{Sarah Masaad\authormark{1,*},  Jakob Declercq\authormark{2}, Stijn Sackesyn\authormark{1}, \\ Ruben Van Assche\authormark{1}, Hasan Salmanian\authormark{1}, Andrzej Polatynski\authormark{3}, Dimitrios Chatzitheocharis\authormark{4,}\authormark{5}, Konstantinos Vyrsokinos\authormark{5}, Tatiana Buriakova\authormark{6}, Christophe Caillaud\authormark{7}, Stylianos Sygletos\authormark{8}, Xin Yin\authormark{2} and Peter Bienstman\authormark{1}}

\address{\authormark{1} Photonics Research Group (PRG), INTEC, Ghent University - imec, 9052 Ghent, Belgium\\
\authormark{2} IDLab, INTEC, Ghent University - imec, 9052 Ghent, Belgium\\
\authormark{3} VPIphotonics GmbH, Berlin, Germany\\
\authormark{4} Now with OneTouch Technology, 9052 Ghent, Belgium\\
\authormark{5} School of Physics, Aristotle
University of Thessaloniki, Thessaloniki, Greece\\
\authormark{6} Ligentec SA, CH1024, Ecublens, Switzerland \\
\authormark{7} III-V lab, a joint lab between Nokia Bell Labs, Thalys Research \& Technology and CEA LETI, Palaiseau, France\\
\authormark{8} Aston Institute for Photonic Technologies (AIPT), Aston University, Birmingham, UK}

\email{\authormark{*}sarah.masaad@ugent.be} 


\begin{abstract*} 
Intensity-modulation direct-detection links must support increasing baudrates and transmission distances while operating under stringent power and cost constraints. However, as data rates and reaches increase, chromatic dispersion induces stronger inter-symbol interference and, after direct detection, frequency-selective fading, thus requiring increasingly powerful equalization. In conventional receivers, this translates into digital equalization whose complexity scales unfavorably with data rate. Photonic-domain equalization offers a hardware-based alternative that operates naturally at line rate and mitigates frequency fading. However, prior demonstrations were not readily adaptable for different rate and/or reach operation. In this paper, we experimentally demonstrate all-optical equalization across 10–46 Gbaud and 10–250 km SSMF in the C-band enabled solely through retraining of the readout layer, achieving up to four orders of magnitude BER improvement over standard DSP equalization. The demonstrator comprises a 16-node spatially multiplexed reservoir, programmable on-chip readout, and co-packaged receiver front-end. To our knowledge, this is the first co-packaged photonic reservoir receiver and the first demonstration of simultaneous baudrate- and reach-flexible equalization using a fixed-topology integrated photonic circuit.\end{abstract*}

\section{Introduction}
The ICT sector currently contributes a carbon footprint comparable to, and in some estimates exceeding, that of global aviation \cite{markuszimmerMoreEmissionsMeet2023}. Thus, as traffic demand continues to grow, system design is increasingly constrained not only by capacity targets but also by energy efficiency and scalability. This constraint is especially relevant in short- and mid-reach optical links, including inter- and intra-data center interconnects, fiber-to-the-home deployments, and mobile front-haul networks. These links are deployed at a large scale, which imposes strict requirements on cost, footprint, and power consumption, in contrast to long-haul systems where complexity and cost can be distributed over longer distances and smaller deployment volumes \cite{chagnonOpticalCommunicationsShort2019, pang200GbpsLane2020}

In these regimes, intensity modulation and direct detection (IM/DD) remains the dominant transceiver architecture due to its relative simplicity and favorable cost–power tradeoffs. However, the scalability and reach of IM/DD links is limited by fiber- and component-induced impairments. In the C-band over standard single-mode fiber (SSMF), chromatic dispersion (CD) results in inter-symbol interference, which after square-law detection induces frequency-selective fading that becomes increasingly severe with with baudrate and transmission distance \cite{chagnonOpticalCommunicationsShort2019}.

Several techniques have been proposed to mitigate frequency selectivity in IM/DD systems and perform signal equalization, including transmitter-based equalization \cite{ran4x256GbpsSilicon2025, 2025OpticalFiber2025}, self-coherent receivers \cite{cheLinearizationDirectDetection2016} like Kramers–Kronig detection \cite{mecozziKramersKronigReceivers2019}, and advanced digital equalization methods such as Volterra equalizers \cite{stojanovicVolterraWienerEqualizers2017, wettlinComplexityReductionVolterra2020} and neural networks \cite{freireNeuralNetworksBasedEqualizers2022}. While effective in specific regimes, these approaches typically introduce additional hardware complexity, bandwidth overhead, stringent component requirements, or substantial DSP load, limiting their suitability for scalable, low-power receivers \cite{zhongDigitalSignalProcessing2018}.

More broadly, continued reliance on increasingly complex DSP faces growing challenges as digital scaling slows, while data rate demands continue to rise rapidly, creating pressure on processing efficiency and power consumption \cite{agrellRoadmapOpticalCommunications2024}. This divergence motivates investigation of alternative processing strategies that reduce dependence on high-rate digital computation.

Integrated photonics provides a promising platform for such approaches. Mature low-loss photonic integration technologies enable the realization of compact photonic processors that offer native operation at line rate, low latency, and parallelism across wavelengths and/or polarizations \cite{shekharRoadmappingNextGeneration2024, mcmahonPhysicsOpticalComputing2023}. Because CD is compensated before the photodetection, photonic equalization can also inherently address the frequency fading problem in IM/DD systems. These properties make shifting the signal conditioning from the digital electronic domain into the analogue photonic domain attractive, thereby relaxing DSP and ADC requirements, and lowering forward-error correction (FEC) overhead while preserving the simplicity of an IM/DD system. 

To ensure solution flexibility and to leverage advances in machine learning (ML), photonic implementations of ML primitives have been extensively investigated \cite{brunnerRoadmapNeuromorphicPhotonics, shastriPhotonicsArtificialIntelligence2021}. Among these approaches, reservoir computing (RC) is particularly well suited for hardware realization because it employs a fixed recurrent dynamical system while requiring training only of a linear readout layer \cite{tanakaRecentAdvancesPhysical2019,yanEmergingOpportunitiesChallenges2024, zhangSurveyReservoirComputing2023}. This structure enables simple training, stable operation, and efficient inference, while naturally exploiting fading memory and high-dimensional state expansion.

Significant progress has been reported in photonic reservoir implementations for signal equalization across a range of platforms and architectures. However, many demonstrated photonic reservoirs rely on bulk or delay-loop configurations that are not readily compatible with compact, integrable, real-time receiver architectures \cite{argyrisPhotonicMachineLearning2018, argyrisPAM4Transmission15502019}. While recent studies have begun to demonstrate hardware-based and online operation \cite{asscheRealtimeAllopticalSignal2025, wangTerabitIntegratedNeuromorphic2025}, implementations integrated directly at the receiver front-end remain limited. Additionally, integrated photonic reservoirs are fabricated with fixed topologies and interconnection delays and are therefore often assumed to operate only within a narrow operating regime. Consequently, prior demonstrations focus on showing single/dual baudrate operation over a restricted reach range or a fixed reach with flexible baudrate operation. However, the ability of a fixed-topology integrated photonic reservoir to generalize across various transmission scenarios in both reach and rate simultaneously has not yet been systematically established.

In this work, we demonstrate a co-packaged, receiver-side photonic reservoir equalizer. Using this integrated device, we experimentally show OOK all-optical signal equalization across a baudrate range from 10 to 46 Gbaud and standard single-mode fiber reaches from 10 km to 250 km in the C-band. Due to limitations of the experimental hardware, the results are extended through simulations of a digital twin up to 100 Gbaud and beyond 500 km. Operation across these regimes is achieved without modification of the reservoir topology and solely through retraining of the readout layer, leveraging the recurrent dynamics and fading memory of the reservoir. The experimental demonstrator comprises a 16-node spatially multiplexed reservoir, an on-chip programmable optical readout based on Mach–Zehnder interferometer heaters, an optical combination tree, and co-packaged optoelectronics including semiconductor optical amplifiers, photodiodes, and transimpedance amplifiers (TIA). These results establish that a fixed integrated photonic reservoir can operate across a broad range of baudrate and reach conditions through readout retraining alone, supporting reuse of a single hardware processor across multiple link configurations. To our knowledge, this is the first demonstration of a co-packaged receiver-side photonic reservoir equalizer, and the first experimental demonstration of flexible operation in both baudrate and reach using a fixed-topology photonic reservoir.

The remainder of this paper is organized as follows. Section 2 introduces reservoir computing and describes the architecture of the reservoir used in this work. Section 3 presents the assembly combining the reservoir and the receiver front-end, together with key design considerations and experimental constraints. Section 4 describes the experimental setup and outlines the training and signal processing procedures. Section 5 presents the experimental results, extends them through simulations to explore operating regimes beyond the experimental hardware limits, and compares the approach with the state of the art. Finally, Section 6 concludes the paper.

\section{Reservoir Computing}
\begin{figure}[H]
\centering\includegraphics[width=\textwidth, trim=1cm 5.9cm 2.5cm 4.7cm,clip]{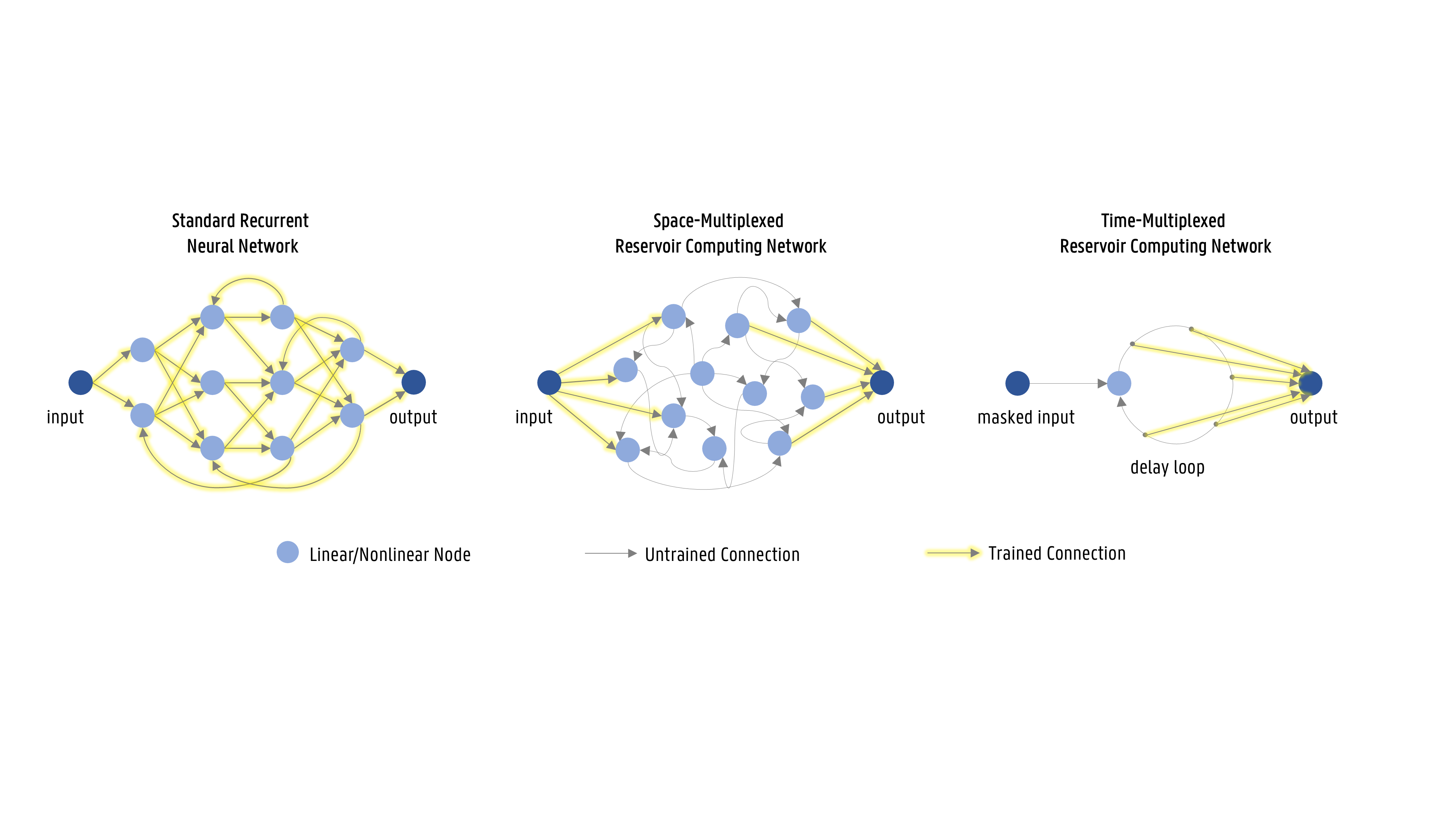}
\caption{Comparison of a recurrent neural network (left) with reservoir computing architectures, including space-multiplexed reservoirs with physically distinct nodes (center) and time-multiplexed delay-based reservoirs using virtual nodes (right). In RC, the internal connections remain fixed while only the readout weights are trained. We follow  visual conventions introduced in \cite{jaurigueConnectingReservoirComputing2022}}
\label{fig:rnn_res}
\end{figure}

Reservoir computing (RC) is a learning-based framework in which a high-dimensional dynamical system (the reservoir) transforms an input time series into a set of internal states, while training is restricted to a linear readout that maps these states to the desired output \cite{lukoseviciusReservoirComputingApproaches2009}. This is contrasted with a standard recurrent neural network (RNN) in Figure \ref{fig:rnn_res}, where all weights are trained compared to the RC's output (and optionally input) weights. 

RC's computational capability arises from fading memory, which embeds recent input history into the current state, and state expansion, which increases dimensionality and can render the target mapping linearly separable at the readout. Because only the readout layer is trained, RC provides a hardware-efficient approach to temporal signal processing with comparatively simple physical realizations and with low training overhead compared to fully trained recurrent neural networks. Photonic RC implements the reservoir in optical hardware, with two common physical realizations being time-delay reservoirs and space-multiplexed reservoirs \cite{vandersandeAdvancesPhotonicReservoir2017}. 

Time-delay reservoirs emulate a multi-node network using a single nonlinear element with delayed feedback, where virtual nodes are created through time multiplexing of a delay loop \cite{appeltantInformationProcessingUsing2011, duportAllopticalReservoirComputing2012, brunnerParallelPhotonicInformation2013} as shown in Figure \ref{fig:rnn_res}. This approach is hardware-efficient in terms of active components but relies on sequential time slicing and long delay paths, which constrain throughput and complicate integration for receiver-front-end operation. In contrast, space-multiplexed reservoirs implement physically distinct nodes connected by optical interconnects, enabling simultaneous state evolution and direct readout evaluation \cite{vandoorneParallelReservoirComputing2011, vandoorneExperimentalDemonstrationReservoir2014} . This parallelism make space-multiplexed architectures more suitable for real-time receiver implementations operating at line rate.

\begin{figure}[htbp]
\centering\includegraphics[width=0.9\textwidth, trim=1.5cm 6.25cm 1.5cm 5.15cm,clip]{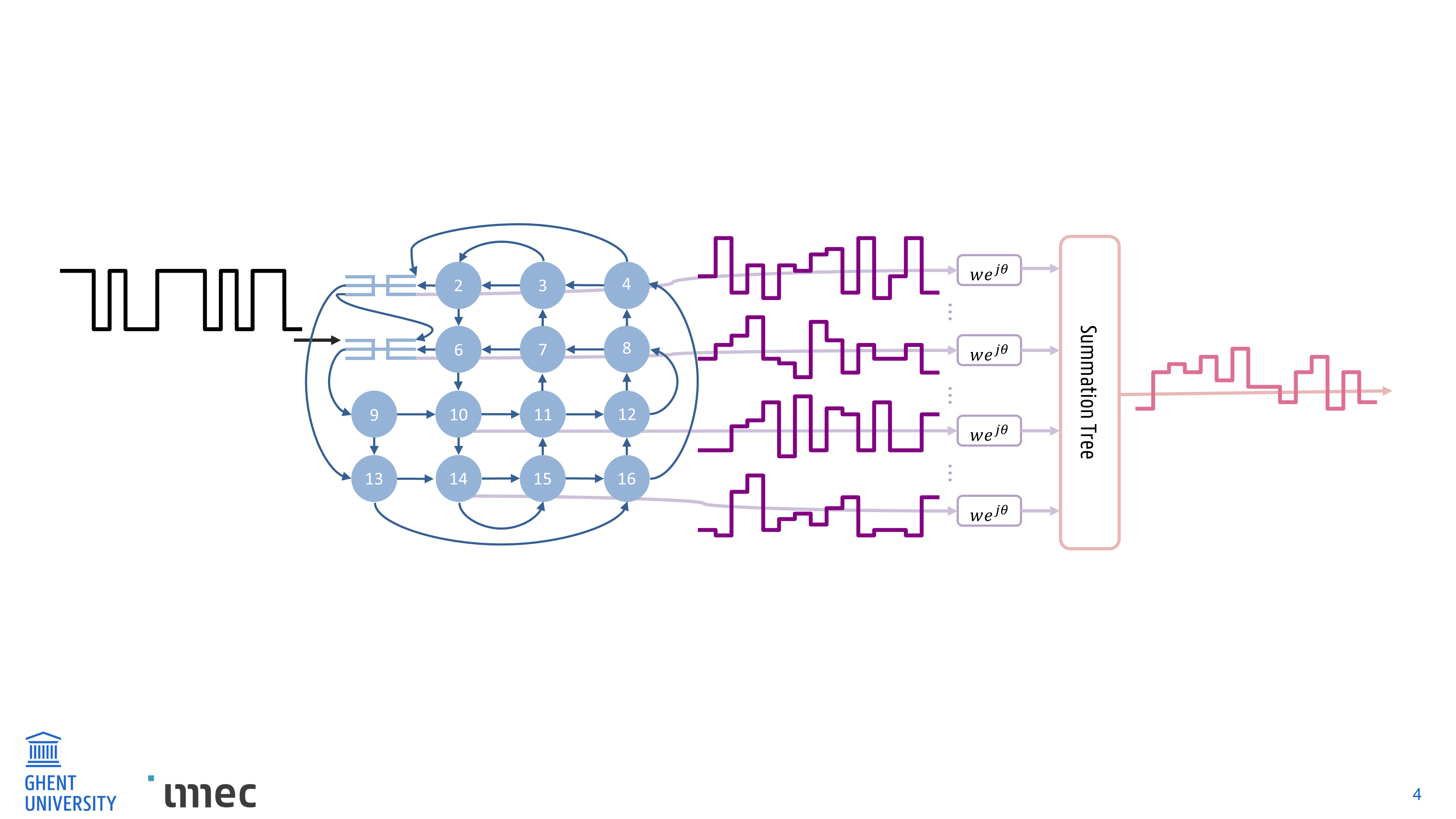}
\caption{Four-port reservoir architecture comprising 16 nodes. Two representative nodes are depicted as MMIs with three inputs and three outputs, while the remaining nodes are abstracted for clarity. Interconnections between nodes define the reservoir topology, with optical weights implemented at the readout. The weighted readout states are combined by a summation tree. The resulting reservoir dynamics are illustrated for an example input signal.}
\label{fig:res_signals}
\end{figure}

Our integrated photonic reservoir implementation is a space-multiplexed reservoir and is referred to as the four-port architecture \cite{sackesynEnhancedArchitectureSilicon2018}. As shown in Fig. \ref{fig:res_signals}, the reservoir employs multimode interferometers (MMIs) as reservoir nodes, where each node receives two inputs from other reservoir nodes and routes two outputs onward within the network. A third input port enables external signal injection and can be selectively used to define reservoir input nodes. A third output port feeds the readout stage, where complex-valued trainable weights are applied. The internal reservoir weights are determined by the fixed waveguide interconnections, which impose phase and, to a lesser extent, amplitude transformations on the propagating optical fields. Fabrication variations, such as sidewall roughness and length variability, introduce device-specific but static weight realizations, resulting in a unique fixed reservoir per fabricated chip.

The interconnecting waveguides are implemented as long spiraled delay lines to introduce controlled temporal delay prior to node mixing, thereby increasing reservoir memory and enabling processing of memory-intensive tasks such as dispersion compensation. For effective operation, delay values are defined relative to the symbol period of the input signal, since the reservoir must provide sufficient memory depth to retain bit-level memory. Prior studies indicate that inter-node delays close to half the symbol period provide favorable performance \cite{katumbaNeuromorphicSiliconPhotonics2019}, and this criterion is commonly used as a baseline design rule. In the present implementation, a distribution of both shorter and longer delay lines is employed rather than a single uniform delay value. This choice was adopted to provide broader delay coverage when evaluating performance under different symbol-rate conditions. Furthermore, several inputs  into the reservoir are used, each with varying delay lengths to enhance the memory dynamics. 

An example of the reservoir operation is illustrated in Figure \ref{fig:res_signals}, where a sequence of input bits is injected into the network. Through the reservoir dynamics, the signal interacts with delayed versions of itself, producing a set of readout states with diverse features. For illustration, the readout is not trained here and the output signal is shown simply as a sum of the reservoir states.

The reservoir is passive and does not require electrical power to operate the nodes or interconnections. The trainable readout weights are implemented optically, with the the present demonstrator using integrated thermo-optic phase shifters. However, the architecture is also compatible with nonvolatile weighting elements, which would further reduce static power consumption once programmed \cite{shekharRoadmappingNextGeneration2024}. In addition, the reservoir itself does not rely on nonlinear optical elements, thereby avoiding the integration complexity and control challenges associated with active nonlinear devices. Instead, the required nonlinearity arises at the photodetector, where square-law detection converts complex-valued optical fields to real-valued electrical intensity signals, providing the nonlinear transformation needed for RC operation \cite{vandoorneExperimentalDemonstrationReservoir2014}. This nonlinearity does not impose an optical power threshold, thus further supporting low-power operation.

\section{Assembly}
\begin{figure}[H]
\centering\includegraphics[width=\textwidth, trim=1.4cm 3cm 3.8cm 0cm, clip]{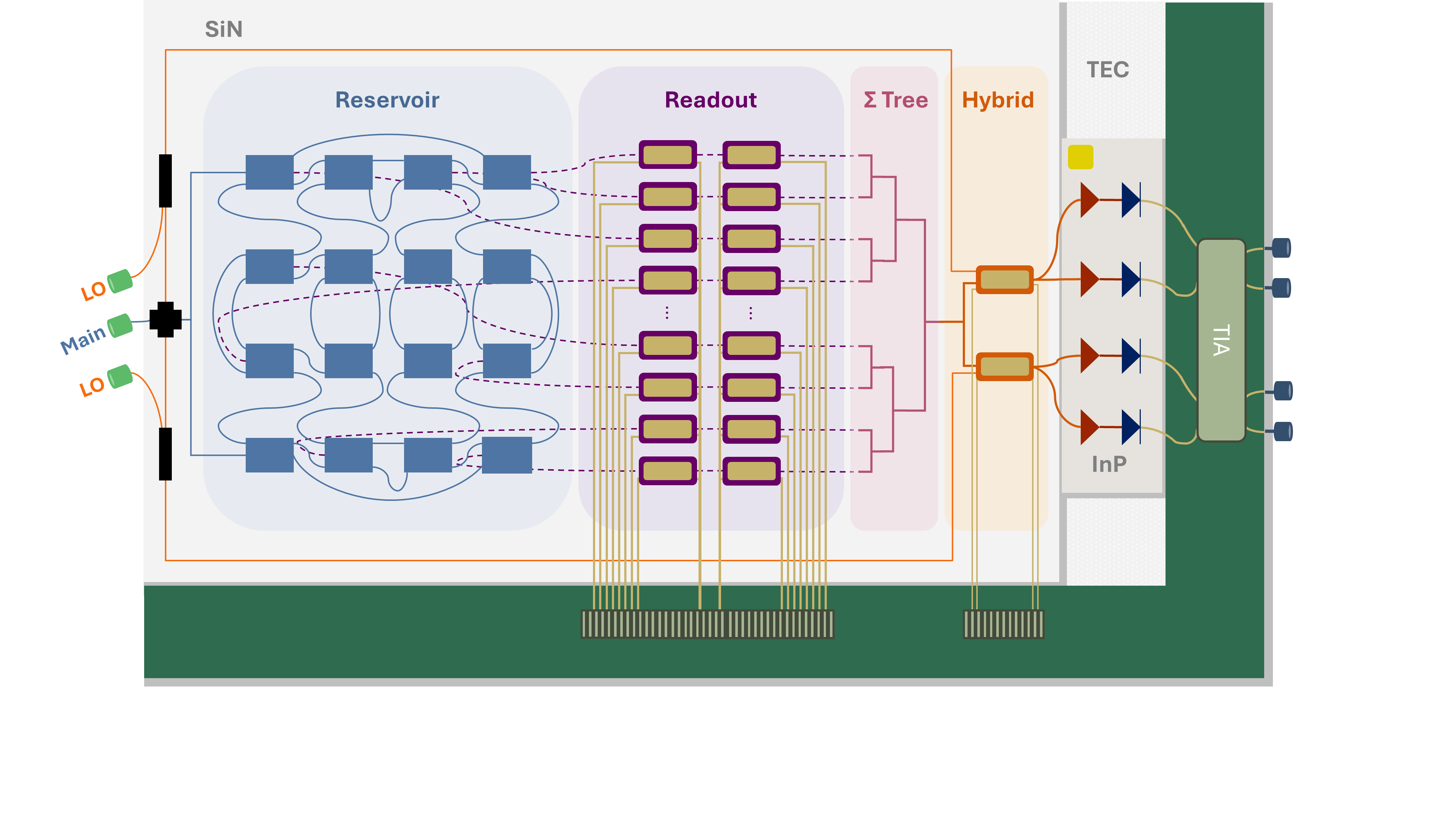}
\centering\includegraphics[width=\textwidth, trim=1.8cm 12.8cm 2.7cm 0.5cm, clip]{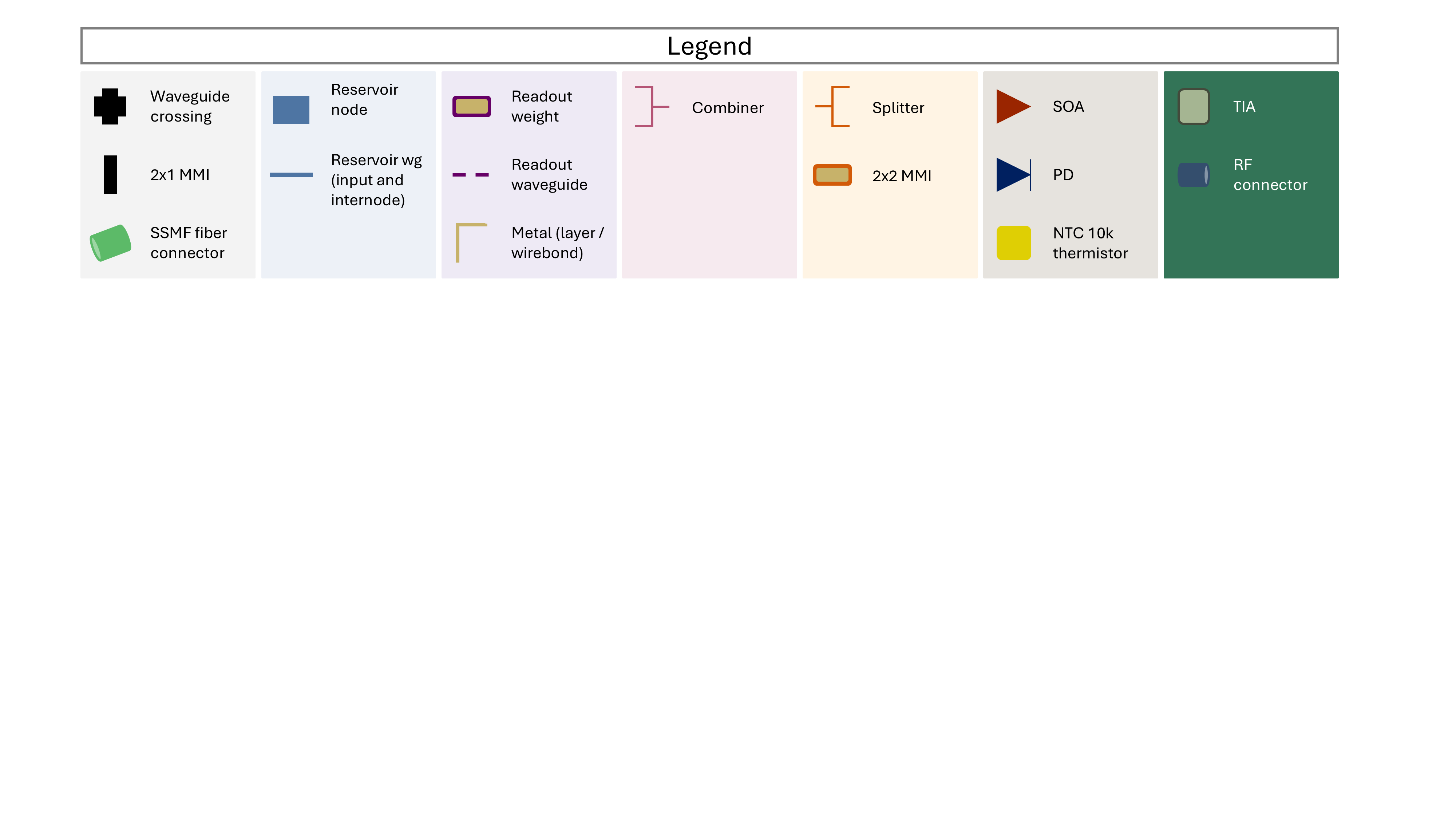}
\caption{System-level schematic of the co-packaged reservoir - receiver demonstrator. The SiN photonic chip comprises an input coupling section with SSMF fiber ports, a recurrent reservoir network, and a readout layer. This is butt-coupled to the SOA-PD chip, where the optical signals are detected. The generated electrical signals are amplified by TIAs and routed to RF connectors. Electrical interfacing also includes the DC control of readout weights. An NTC (Negative Temperature Coefficient) thermistor and TEC (Thermoelectric Cooler) block are included for thermal monitoring and stabilization. The legend summarizes the symbols used.}
\label{fig:assembly}
\end{figure}

\begin{figure}[H]
\centering\includegraphics[width=0.5\textwidth, trim=0cm 0cm 0cm 0cm,clip]{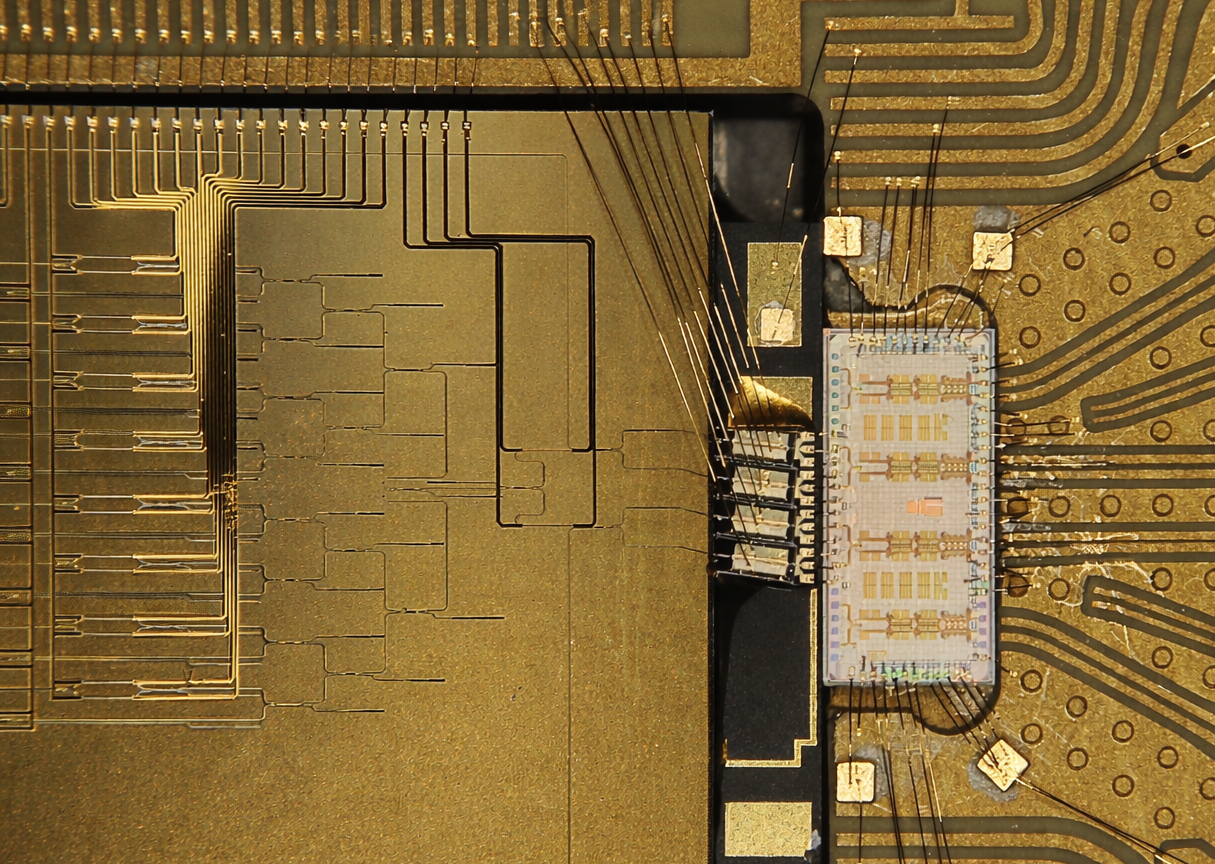}
\caption{Microscopic image showing co-integration of three chips: the SiN chip (left), InP chip hosting the SOA-UTC photodiodes (center) and the BiCMOS chip hosting the TIAs (right). Note that the reservoir itself is outside the frame of this image, where only a portion of the readout, the summation tree and hybdird are shown on the SiN chip.}
\label{fig:chip2}
\end{figure}

The reservoir equalizer was integrated into a full receiver assembly developed within the EU Horizon 2020 NEBULA project \cite{nebula871658}, as illustrated in Fig. \ref{fig:assembly}. This assembly combines a photonic reservoir computing chip with an integrated coherent receiver front-end. The reservoir chip, fabricated on the Ligentec SiN platform \cite{ligentecLigentecSiliconNitride}, implements the recurrent photonic network used for equalization. This SiN chip is butt-coupled to an InP chip from III-V lab \cite{IIIVLab} that monolithically integrates semiconductor optical amplifiers (SOA) and uni-travelling carrier (UTC) photodiodes, enabling high-speed optoelectronic conversion with integrated optical gain. The electrical frontend consists of a linear four-channel TIA chip fabricated in a 55~nm BiCMOS technology, providing broadband, low-noise amplification of the detected signals. Optical packaging of the system was performed by PHIX \cite{phix}, including epoxied edge-coupled fibers and photonic chip alignment. A microscopic image of the three co-integrated chips is shown in Figure \ref{fig:chip2}.  

The receiver frontend can also be configured for single-ended direct detection by enabling only one SOA branch. Although this configuration introduces additional loss compared to a dedicated direct-detection receiver, it provides architectural flexibility and allows the same hardware platform to support both IM/DD and coherent measurement scenarios. In the present work, the assembly is operated in IM/DD mode.

The SiN chip hosts a 16-node reservoir, as described in Section 2. The optical input signal is injected through the middle of the three fibers shown in Fig. \ref{fig:assembly} and is distributed to the reservoir input nodes via on-chip splitter networks. The node outputs are routed to the readout stage through dedicated waveguides, where the readout weights are implemented using a Mach–Zehnder interferometer with two phase shifters to control the amplitude and phase, enabling complex-valued weighting. The weighted channels are then combined using a hierarchical summation tree, reducing the multi-node states to a single optical output that is forwarded to the receiver section. 

\begin{figure}[htbp]
\centering\includegraphics[width=0.5\textwidth, trim=0cm 9.4cm 17cm 0.4cm,clip]{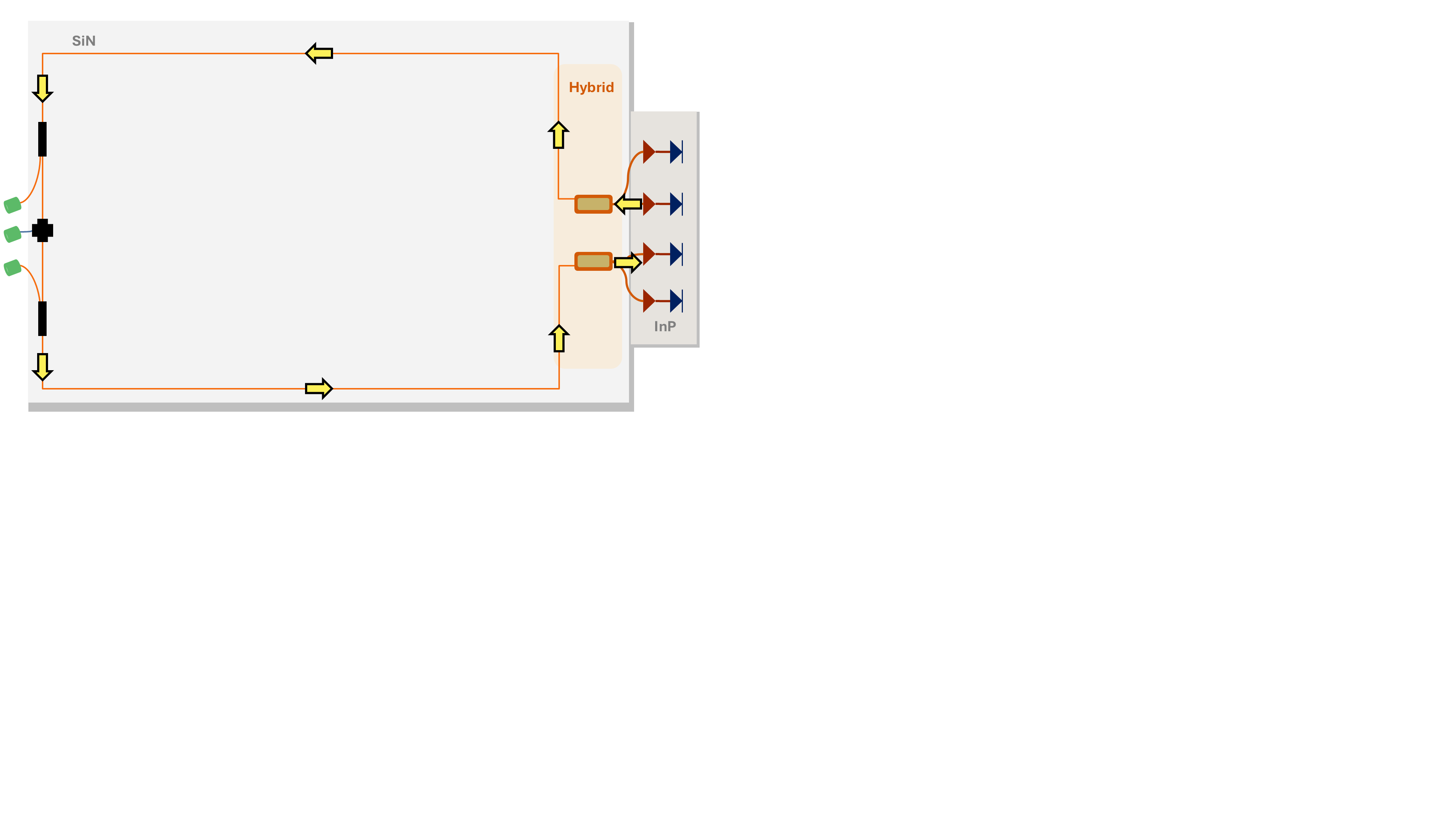}
\caption{LO routing and alignment scheme in the assembly, with irrelevant details omitted for clarity. The LO is injected through two fiber pigtails feeding the hybrid MMIs, with interconnected paths facilitating alignment between the SiN reservoir chip and the InP receiver chip. During alignment, the upper SOA is biased so that its ASE couples from the InP chip to the SiN chip, propagates along the LO waveguides, and couples back to the InP chip, where detection confirms correct chip-to-chip alignment. A waveguide crossing is introduced where the LO paths intersect the reservoir input.}
\label{fig:signal_flow1}
\end{figure}

The receiver subsection consists of a 90° optical hybrid followed by four SOA–photodiode branches, forming a coherent detection structure. The local oscillator (LO) is injected into the hybrid through two separate fiber pigtails, each feeding one of the hybrid's MMIs, as shown in Fig.  \ref{fig:signal_flow1}. To facilitate the relative alignment of the SiN chip and the InP chip, the two LO paths are also connected, thus serving as a direct optical feedback during alignment. Alignment is then verified by biasing the upper SOAs to generate amplified spontaneous emission (ASE) and maximizing the detected power at the corresponding lower photodiodes. To realize this routing, a waveguide crossing had to be introduced making this path cross with the reservoir input, as indicated in Fig. \ref{fig:signal_flow1}.

\subsection {Assembly Constraints}
\begin{figure}[htbp]
\centering\includegraphics[width=0.5\textwidth, trim=0cm 9.6cm 17cm 0.7cm,clip]{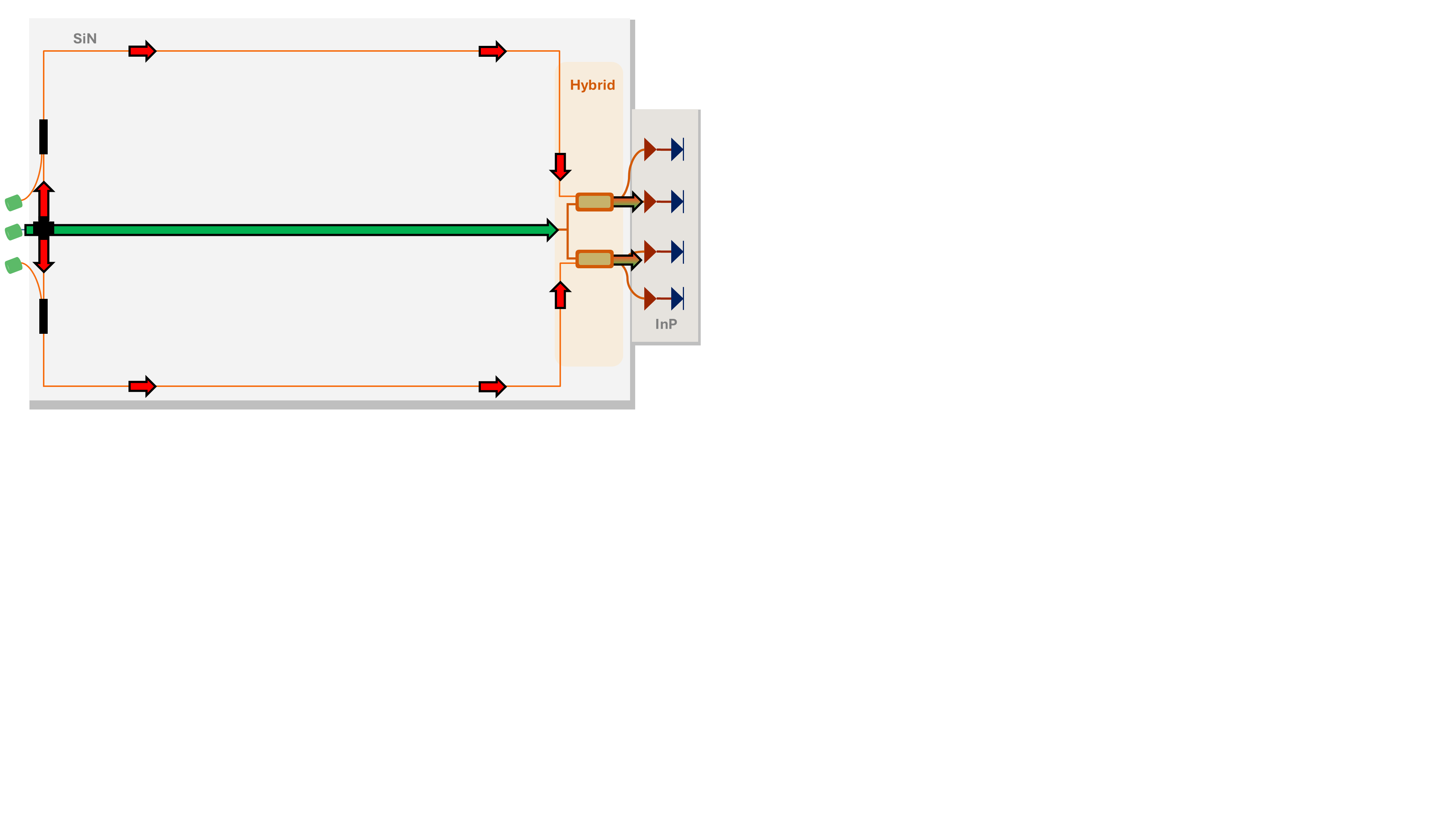}
\caption{Crosstalk originating at the crossing allows a fraction of the signal (indicated in red) to bypass the reservoir equalization path (indicated in green) and couple directly toward the hybrid. Although the crosstalk at the crossing is ~20 dB below the main signal, the significantly higher insertion loss of the reservoir path results in comparable power levels at the hybrid, introducing distortion and limiting the achievable equalization gain.}
\label{fig:crosstalk}
\end{figure}

Two practical limitations arose during operation. First, the waveguide crossing used to realize the joining of the LO paths introduced parasitic optical crosstalk between the reservoir input path and the LO paths, as shown in Figure \ref{fig:crosstalk}. As a result, a fraction of the modulated signal can bypass the reservoir network and reach the hybrid directly through the LO waveguides. This crosstalk contribution was measured to be on the order of 20 dB below the main signal at the crossing itself. However, the effective impact is still significant given that the reservoir path experiences higher insertion loss due to the large number of components, weighting elements, summation tree, and overall interference in the reservoir. In contrast, the LO path avoids all of these components. Since the bypassed signal does not benefit from reservoir equalization, it degrades the achievable system performance and limits the maximum equalization gain that can be observed experimentally. 

\begin{figure}[H]
\centering\includegraphics[width=0.95\textwidth, trim=0cm 0.6cm 2cm 5.2cm,clip]{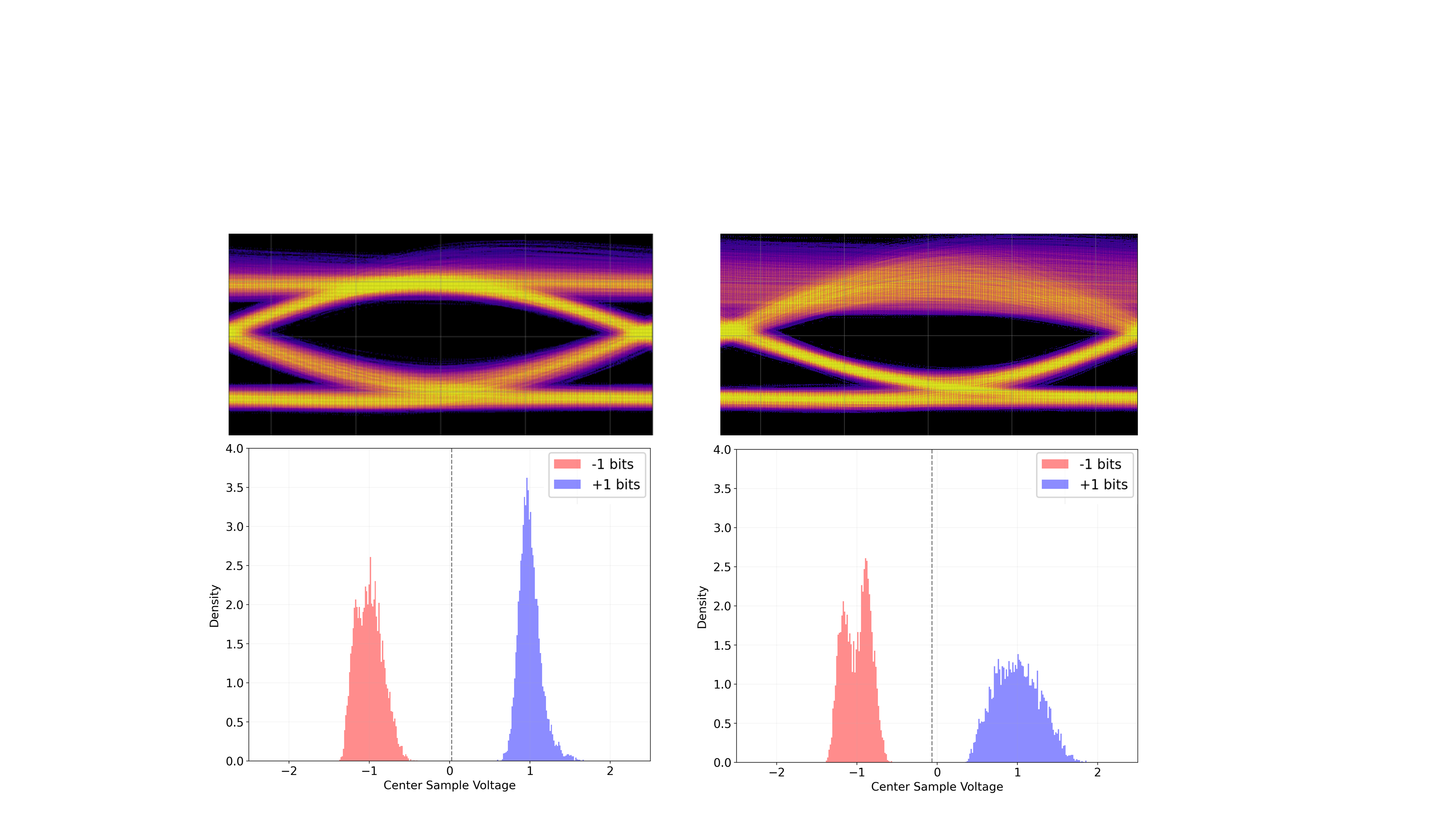}
\caption{ Eye diagrams and histograms of received signal after SOA damage. Left:  0~dBm input power. Slight patterning in the -1 bits can be seen from the histogram. Right: 4~dBm input power. More pronounced patterning is seen in the -1 bits. However, spread due to noise is reduced. +1 bits show loss of a well-defined level cluster}
\label{fig:SOA_damage}
\end{figure}

Second, degradation of the assembly SOAs was observed during operation, consistent with unintended reverse-bias stress leading to altered carrier dynamics and increased susceptibility to gain saturation. After this event, the SOAs exhibited reduced small-signal gain, increased nonlinear distortion, and some pattern-dependent effects. These impairments are reflected in the measured responses in Figure \ref{fig:SOA_damage}.

Despite these hardware constraints and nonideal conditions, the reservoir was still successfully trained and deployed. Indeed, the experimental results presented later in this work demonstrate that the integrated photonic reservoir maintains effective equalization capability across a wide range of transmission scenarios, highlighting the practical robustness of the architecture and its tolerance to hardware imperfections encountered in a real photonic implementation. 

\section{Experimental Setup and Processing Pipeline}

\begin{figure}[htbp]
\centering\includegraphics[width=\textwidth, trim=0.2cm 5cm 8.65cm 5.8cm,clip]{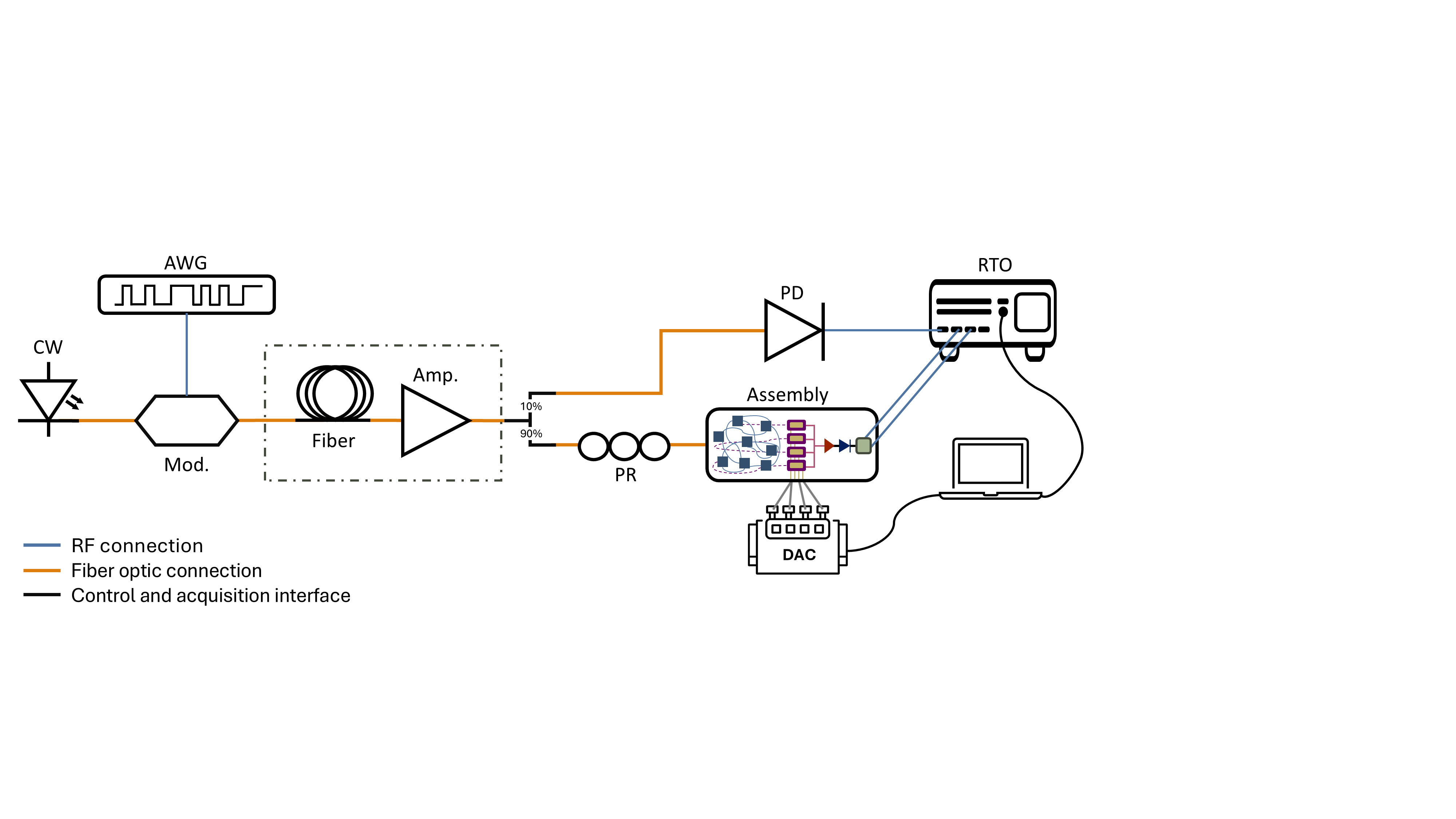}
\caption{Experimental setup. CW: Continuous Wave Laser, AWG: Arbitrary Waveform Generator, Mod.: Modulator, Amp.: Amplifier, PR: Polarization Rotator, PD: Photodiode, DAC: Digital-to-Analogue Converter, RTO: Real-Time Oscilloscope}
\label{fig:setup}
\end{figure}

The experimental setup is shown in Fig. \ref{fig:setup}. A continuous-wave laser at 1550 nm was used as the optical source and was modulated using an external modulator driven by an arbitrary waveform generator (AWG). The AWG was operated at 64 GSa/s and programmed with 15,000 NRZ symbols generated using a cryptographically secure pseudorandom number generator provided by Python’s secrets module. Modulation baudrates between 10 and 46 Gbaud were evaluated. For baudrates up to 32 Gbaud, rectangular pulses with a transition time of 0.7 the symbol period were used. At higher baudrates, pulse shaping was applied due to the limited AWG sampling rate. In these cases, Nyquist-shaped pulses were generated using a raised-cosine filter with roll-off factor of 0.7.

Transmission distances ranging from 10 km to 250 km of standard single-mode fiber were tested using fiber spans with distributed inline amplification. Optical amplifiers were inserted as required to maintain approximately 21 dBm output power at the final amplification stage, while ensuring that the launch power into each fiber span remained at or below 0 dBm.

At the receiver side, the optical signal was split using a 90/10 coupler, with 90\% directed to the reservoir assembly and 10\% to a reference 50 GHz benchtop photodiode. Around 10 dBm was received in the reference path, which is at the higher end of the PD's linear regime, and was used for DSP benchmarking via offline T/2-spaced feed-forward equalization (FFE). The reservoir path included a polarization controller to align the input state before coupling into the reservoir fiber pigtail. Within the assembly, only a single SOA was biased to allow direct detection behaviour. For both the reservoir and reference paths, the detected electrical signals were digitized using a real-time oscilloscope operating at 256 GSa/s.

Equalization in the reservoir is realized by determining a set of readout weights that minimize detection errors at the output. These weights are obtained through an on-chip learning procedure implemented using hardware-in-the-loop optimization. This approach bypasses the need to probe the internal weights and compute them through a digital twin, or to rely on models that account for crosstalk and other hardware imperfections. Instead, the network is treated as a black box in which tunable elements define the system behavior.

A range of training approaches have been developed for such physical networks, including gradient-based and gradient-free optimizations \cite{momeniTrainingPhysicalNeural2025}. In this work, readout training is performed using the covariance matrix adaptation evolution strategy (CMA-ES) \cite{hansenCMAEvolutionStrategy2006}, a population-based, gradient-free optimization algorithm well suited to noisy objective functions. CMA-ES iteratively proposes candidate weight vectors based on the statistical distribution of previously successful solutions, progressively refining the search region until convergence. This means that in our setup, candidate weight vectors are optimized directly based on measured system performance.

\begin{figure}[H]
\centering\includegraphics[width=\textwidth, trim=3cm 3.3cm 3cm 1.4cm,clip]{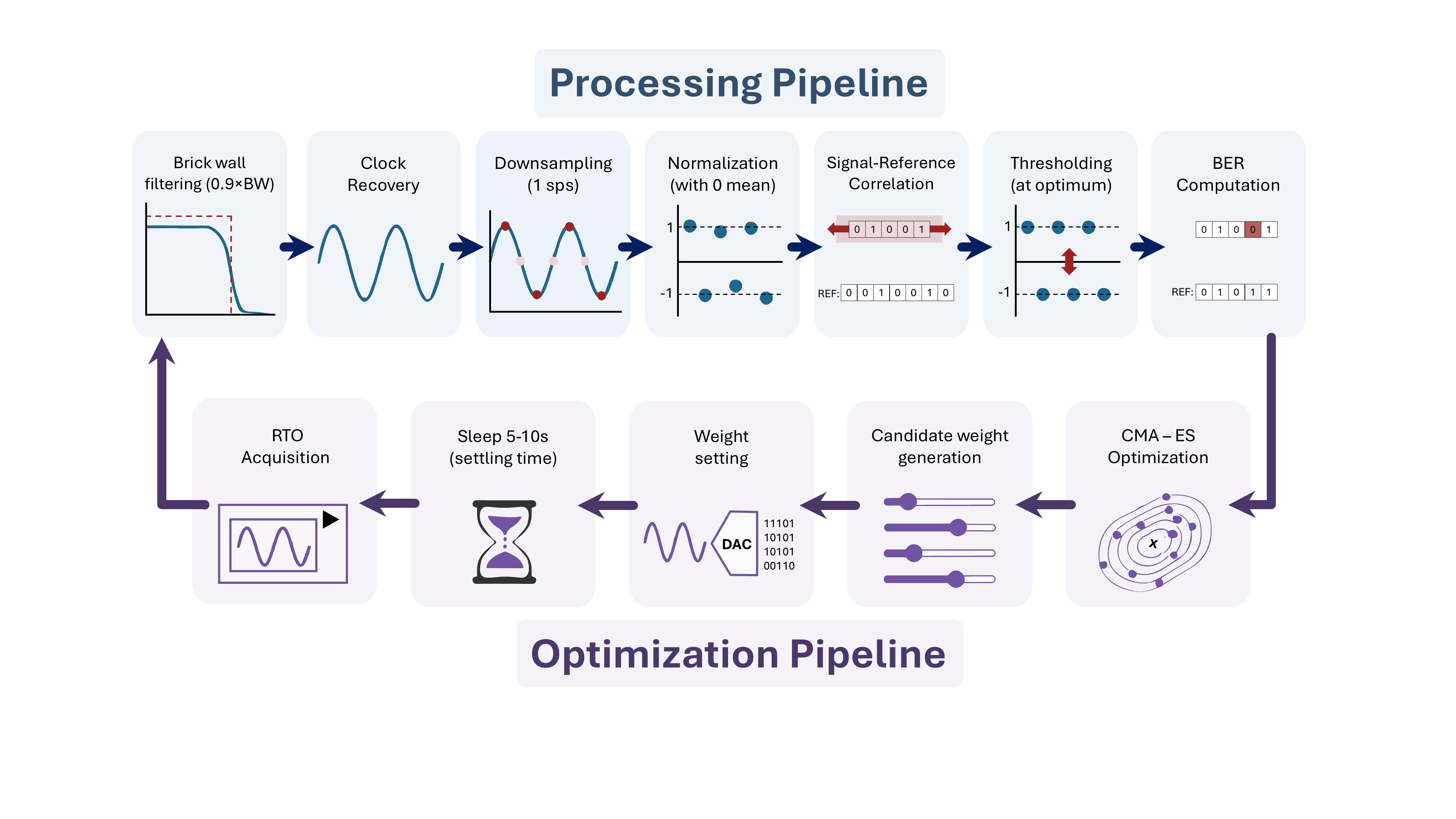}
\caption{Hardware-in-the-loop CMA-ES optimization and signal processing pipeline. Candidate readout weights proposed by the optimizer are programmed on the chip. Then, the resulting signal is acquired and processed to compute the BER, and the measured performance is fed back to update the next generation of candidate weights. The processing pipeline includes filtering, clock recovery, resampling, normalization, thresholding, synchronization, and BER evaluation.}
\label{fig:processing_optimization}
\end{figure}

To apply CMA-ES in the experimental setup, a closed optimization loop is established between the optimizer and the photonic hardware. Each set of weights proposed by the optimizer is programmed onto the chip. Then, the resulting equalized signal is acquired and processed (using the pipeline described in the next paragraph), and the corresponding bit error rate (BER) is computed as the objective metric. This measured BER is returned to the optimizer, which updates its candidate distribution and generates a new population of weight vectors for the next iteration.

The evaluation processing pipeline mentioned above is summarized in Fig. \ref{fig:processing_optimization}. The waveform acquired from the real-time oscilloscope is first filtered and resampled, followed by clock recovery to determine the optimal sampling phase. The signal is then down-sampled to the symbol rate, normalized, and thresholded to produce a binary sequence. The decision threshold is optimized for each signal to maximize detection performance. Bit synchronization is then applied prior to BER computation against the known transmitted pattern. 

Figure \ref{fig:processing_optimization} also illustrates the CMA-ES optimization loop described above. In addition to the standard optimizer steps, the loop also includes a short settling period to allow the weights to stabilize and thermal equilibrium to be reached. This step is necessary when operating physical hardware. Additional constraints are imposed on the optimization (not shown in the figure), including bounding the weights, limiting the current range to a single full MZI sweep, and enforcing a resolution of 1 mA. To ensure that the BER is representative of the performance of the set of weights, the average BER over five acquired signals is used, reducing the impact of stochastic variability and measurement noise. Furthermore, during training, the BER may reach zero, preventing further improvement using BER as the optimization objective. In such cases, a single-sided mean-square error (MSE) objective is used, where the samples are clipped at 1 and -1 before MSE is calculated. This is done such that only deviations toward the decision boundary are penalized, providing a more robust optimization metric under experimental noise. 

This training procedure is repeated for each transmission scenario. Once the optimal readout weights have been identified, the optimization loop is terminated and the evaluation phase begins. During evaluation, the optimized weights are programmed on the chip and the test data are processed using the same signal-processing pipeline shown in Fig.~\ref{fig:processing_optimization}. The only difference is that the decision threshold is not re-optimized on the test data. Instead, the threshold determined from the training dataset is retained to ensure unbiased performance assessment.

To provide a baseline for benchmarking the reservoir equalizer, a DSP-based equalizer was implemented. In this configuration, the optical reservoir processor was replaced by a 16-tap fractionally spaced feed-forward equalizer (FFE) operating at two samples per symbol. The equalizer coefficients were obtained using linear regression to minimize the error with respect to the known target sequence. The same signal-processing pipeline shown in Fig.\ref{fig:processing_optimization} was applied to the detected waveform. After training, the optimized coefficients were applied to the test dataset, and the decision threshold determined from the training data was held fixed during evaluation.

\section{Results and Discussion}
\subsection{Experimental Results}
\begin{figure}[H]
\centering\includegraphics[width=0.89\textwidth, trim=0cm 1.0cm 0cm 0.7cm,clip]{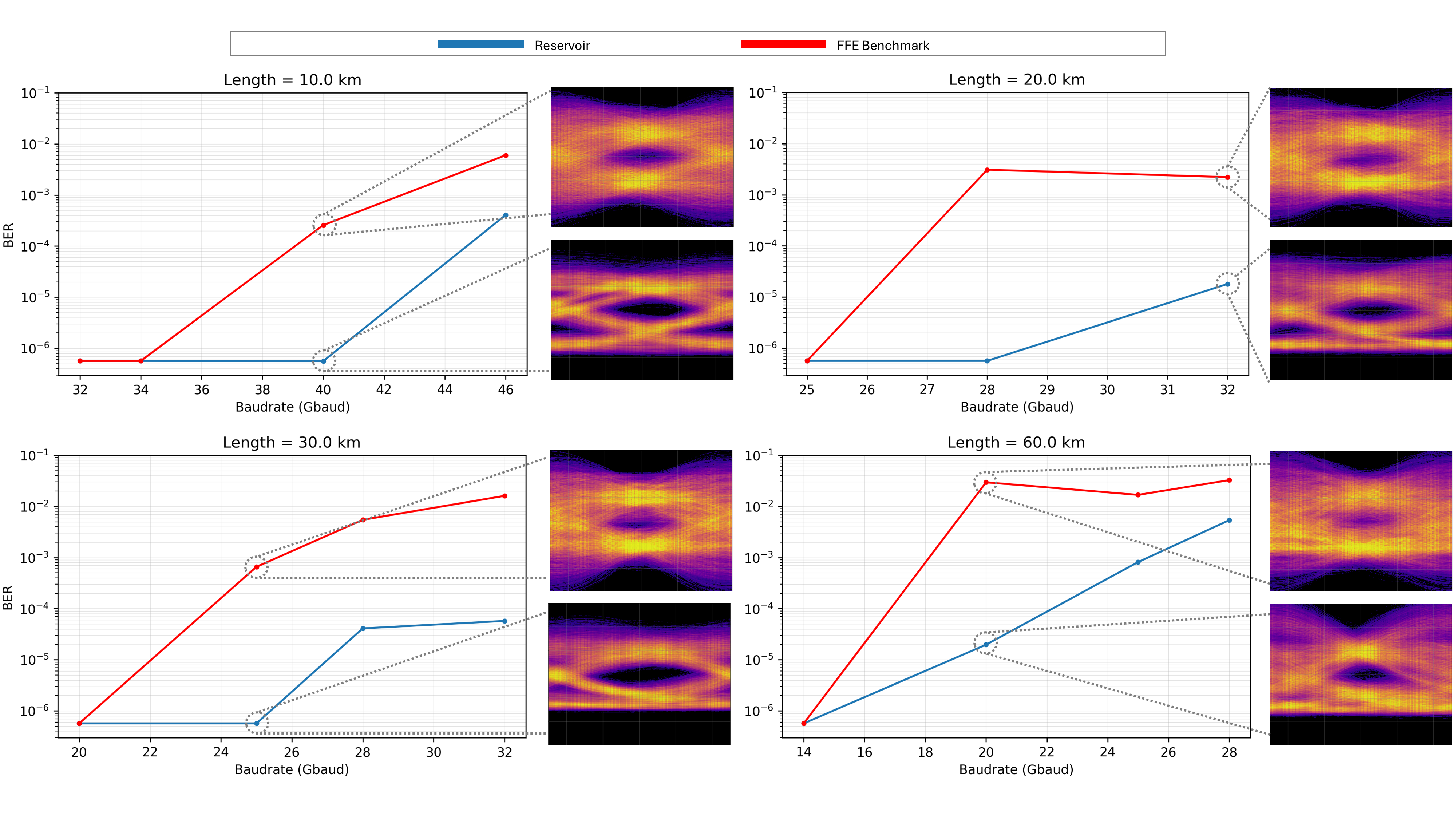}
\caption{BER results grouped by fiber length (10, 20, 30, and 60 km). For each length, multiple baudrates are evaluated as indicated on the x-axis, with BER reported on the y-axis. Reservoir equalization (shown in blue) is compared against the DSP benchmark obtained using the benchtop PD (red).  Side panels show representative eye diagrams after equalization for both the reservoir and benchmark at selected operating points}
\label{fig:baud_results}

\end{figure}

\begin{figure}[H]
\centering\includegraphics[width=0.89\textwidth, trim=0cm 9.4cm 0cm 0.7cm,clip]{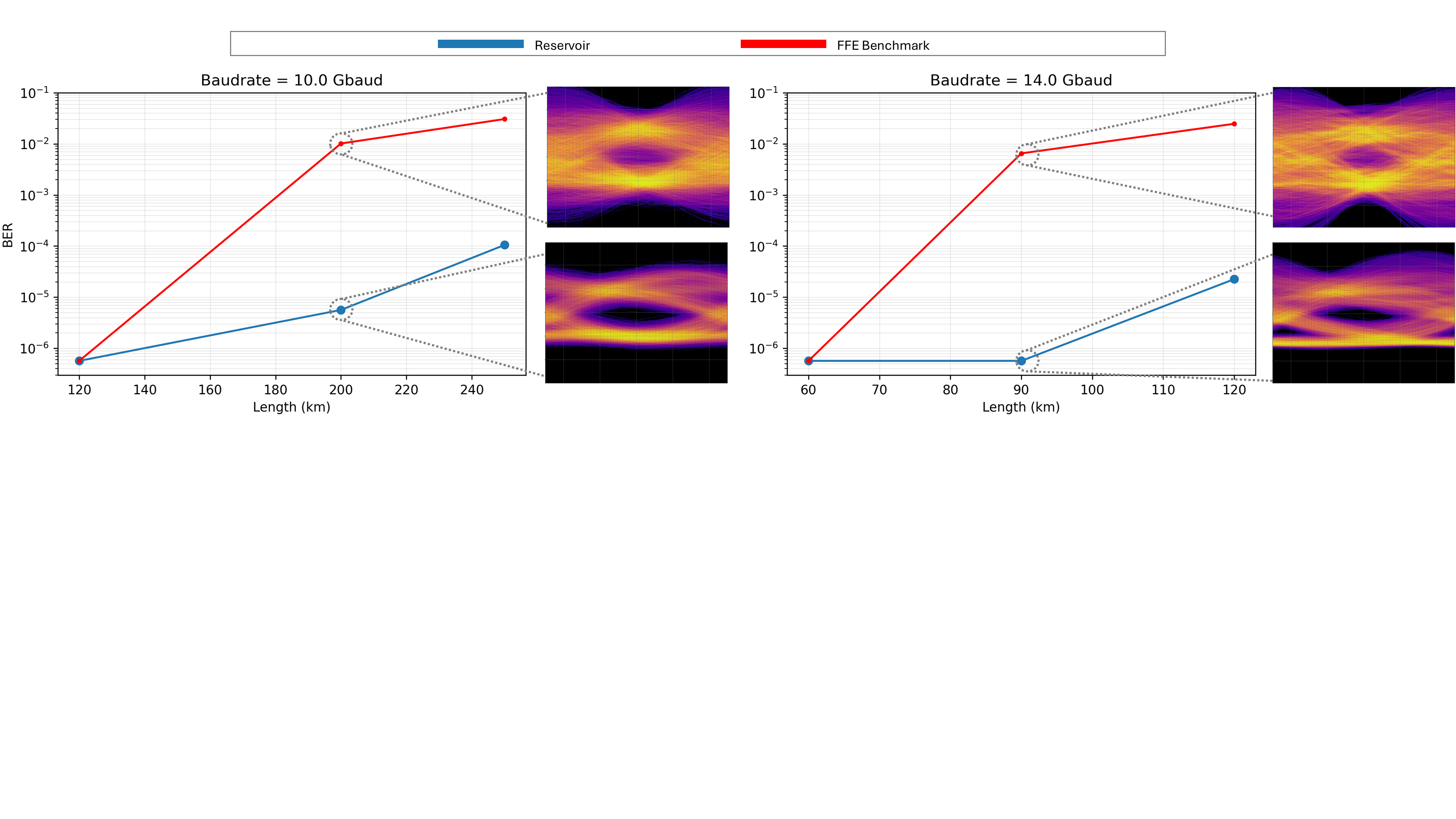}

\caption{BER results grouped by baudrate  (10, 14 Gbaud), with fiber length shown on the x-axis and BER on the y-axis. Measurements extend up to 250 km for 10 Gbaud and up to 120 km for 14 Gbaud. Reservoir equalization (shown in blue) is compared against the DSP benchmark obtained using the benchtop PD (red). Side panels show representative post-equalization eye diagrams for both reservoir and benchmark cases at select data points.}
\label{fig:length_results}

\end{figure}

The experimental results evaluate the reservoir across a wide range of transmission conditions by varying both baud rate and fiber length. Each data point is computed over a test set exceeding 1.7 million bits. The resulting bit error rates (BER) are presented in Figs. \ref{fig:baud_results} and \ref{fig:length_results}, together with the DSP benchmark. The benchmark is obtained from the 10\% reference receiver top branch shown in Fig. \ref{fig:setup}, corresponding to a benchtop photodiode operated within its optimal power range. This reference dataset was acquired concurrently with the reservoir measurements and processed offline using a T/2 16-tap feed-forward equalizer (FFE). The lowest reported BER is the resolution limit, corresponding to the inverse of the test dataset. 

In Fig. \ref{fig:baud_results}, the results are grouped by baud rate for fixed transmission distances. For shorter fiber lengths, higher baud rates are included. For example, the first subplot shows equalization at 32, 34, 40, and 46 Gbaud over 10 km. At 32 and 34 Gbaud, corresponding to weaker dispersion, the FFE performs comparably to the reservoir, consistent with the limited number of frequency nulls. As the baud rate increases and larger dispersion is accumulated, the reservoir achieves one to two orders of magnitude lower BER than the FFE. Due to limitations in the AWG sampling rate and modulator bandwidth, the back-to-back signal quality progressively degrades beyond 32 Gbaud, preventing reliable evaluation above the 46 Gbaud result reported. These higher-rate conditions will therefore further be explored through simulations in the following subsection.

Figure \ref{fig:baud_results} also includes results at longer transmission distances. For example, at 60 km, baud rates of 14, 20, 25, and 28 Gbaud are evaluated, where the reservoir consistently provides multi-order-of-magnitude BER reduction relative to the DSP baseline. 

At lower baud rates (10–14 Gbaud), significantly longer transmission distances can be supported. These conditions are therefore further explored in Fig. \ref{fig:length_results}, where the reach is extended up to 250 km. To the best of our knowledge, this represents the longest experimentally demonstrated transmission distance compensated using a photonic reservoir equalizer. The results show the reservoir maintains strong performance, achieving approximately two to four orders of magnitude BER improvement over the FFE.

Overall, across transmission distances from 10 km to 250 km and baud rates from 10 to 46 Gbaud, the reservoir consistently outperforms the DSP baseline by multiple orders of magnitude. Notably, this performance advantage is achieved with a single, hardware-configured equalizer, in contrast to the baud-rate-specific architecture required for the FFE. 

The experimental performance is partially limited by non-idealities, including by parasitic crosstalk through the local oscillator routing path described in Fig. \ref{fig:crosstalk}, which introduces a residual unequilibrated signal component, as well as by bandwidth constraints of the available instrumentation. As we will show in the following subsection, simulations indicate that removing these mitigable limitations enables further extension of both baud rate and transmission reach.

\subsection{Simulation Results}

To further assess the capabilities of the reservoir architecture, a photonic circuit model was simulated using VPIphotonics Design Suite \cite{VPITransmissionMaker} that reproduces the reservoir's physical characteristics. In particular:

\begin{itemize}[nosep]

    \item Waveguide crossings at the reservoir input were modeled with propagation delay, crosstalk, and insertion loss.
    \item MMIs were modeled with propagation delay and insertion losses drawn from a statistical distribution.
    \item All waveguides were modeled with their exact physical lengths, including the associated propagation delays, together with an additional random phase assigned to each waveguide.
    \item Additional losses arising from waveguide crossings within the reservoir and bends introduced by spiral waveguides were included as extra interconnection losses.
    \item The readout layer was modeled with propagation delay and a random distribution of MMI and MZI losses, together with a finite MZI extinction ratio.
\end{itemize}

The numerical values of the simulation and statistical distributions were derived from the process design kit (PDK) and reflect the expected on-chip behavior of the components. This modeling approach captures the key physical effects governing reservoir operation, including propagation delays, relative path differences, insertion losses, and crosstalk. In order to isolate the intrinsic performance of the reservoir itself, we did not include the parasitic crosstalk path in the readout summation, nor the nonlinearity of the SOA.

\begin{figure}[htbp]
\centering\includegraphics[width=0.6\textwidth, trim=4cm 5.5cm 12cm 8.5cm,clip]{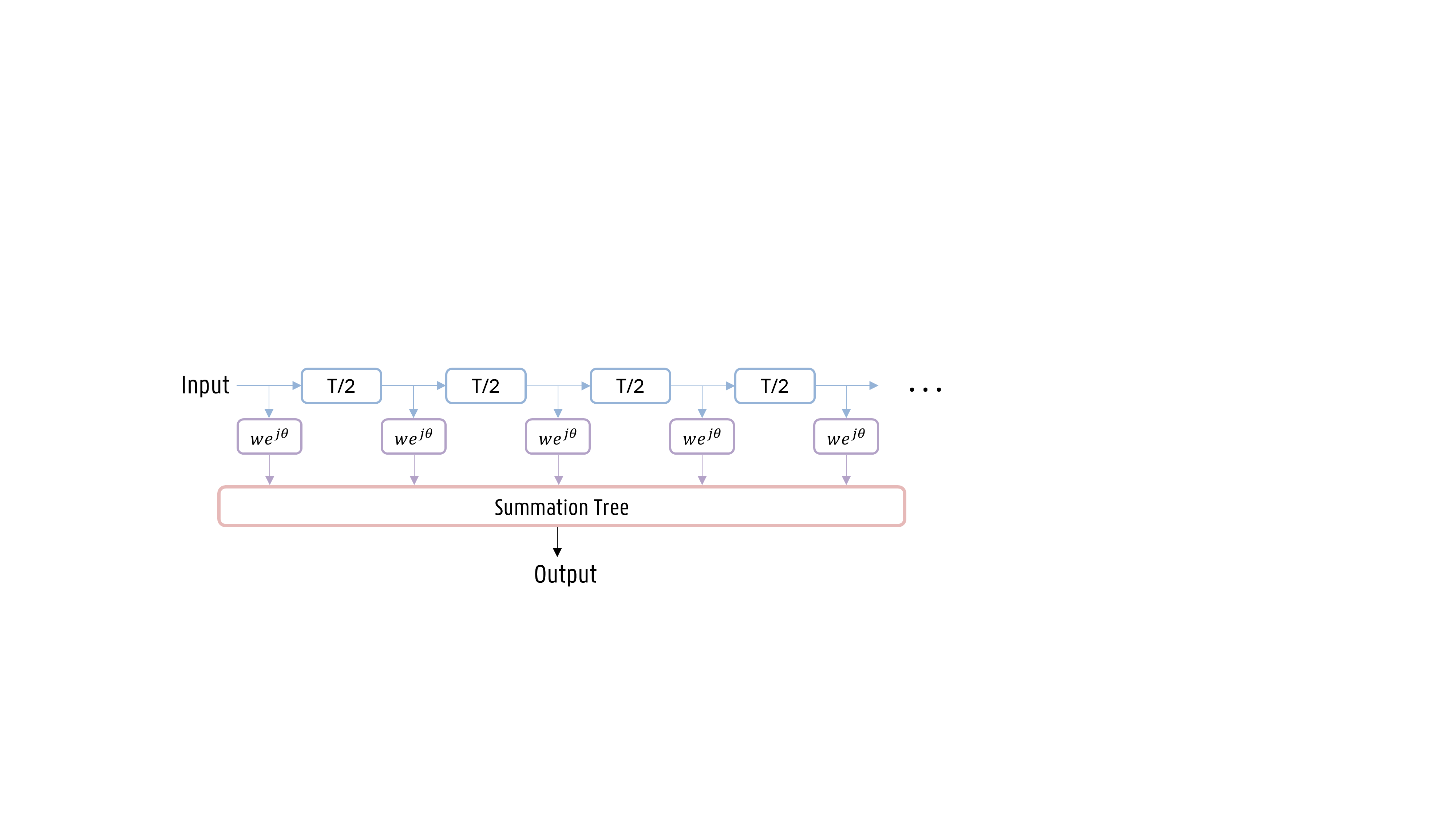}

\caption{Schematic of the optical tapped delay line baseline architecture, implemented as a cascade of 1×2 MMIs and waveguides. One output of each stage contributes to the readout, while the other propagates forward, forming a 16-stage delay line.}
\label{fig:tdl_schematic}

\end{figure}

To benchmark the performance of the proposed architecture against an alternative optical equalization approach, an optical tapped delay line (OTDL) was simulated as a reference. This comparison highlights the advantages of the reservoir in terms of programmability and versatility, particularly in its ability to exploit recurrent dynamics for enhanced memory and equalization performance. The OTDL was implemented using cascaded 1×2 MMIs connected by waveguides, as shown in Fig. \ref{fig:tdl_schematic}. As this architecture serves only as a baseline, equal MMI losses were assumed, and the only modeled imperfections were random waveguide phases. A total of 16 stages were used, where in each stage one MMI output feeds the readout while the other propagates to the next stage. The interconnecting waveguides introduce a delay corresponding to half a symbol period. As multiple baud rates are evaluated, a middle value corresponding to 40 Gbaud was used to define this delay.

\begin{figure}[H]
\centering\includegraphics[width=0.6\textwidth, trim=0cm 8cm 14cm 6.7cm,clip]{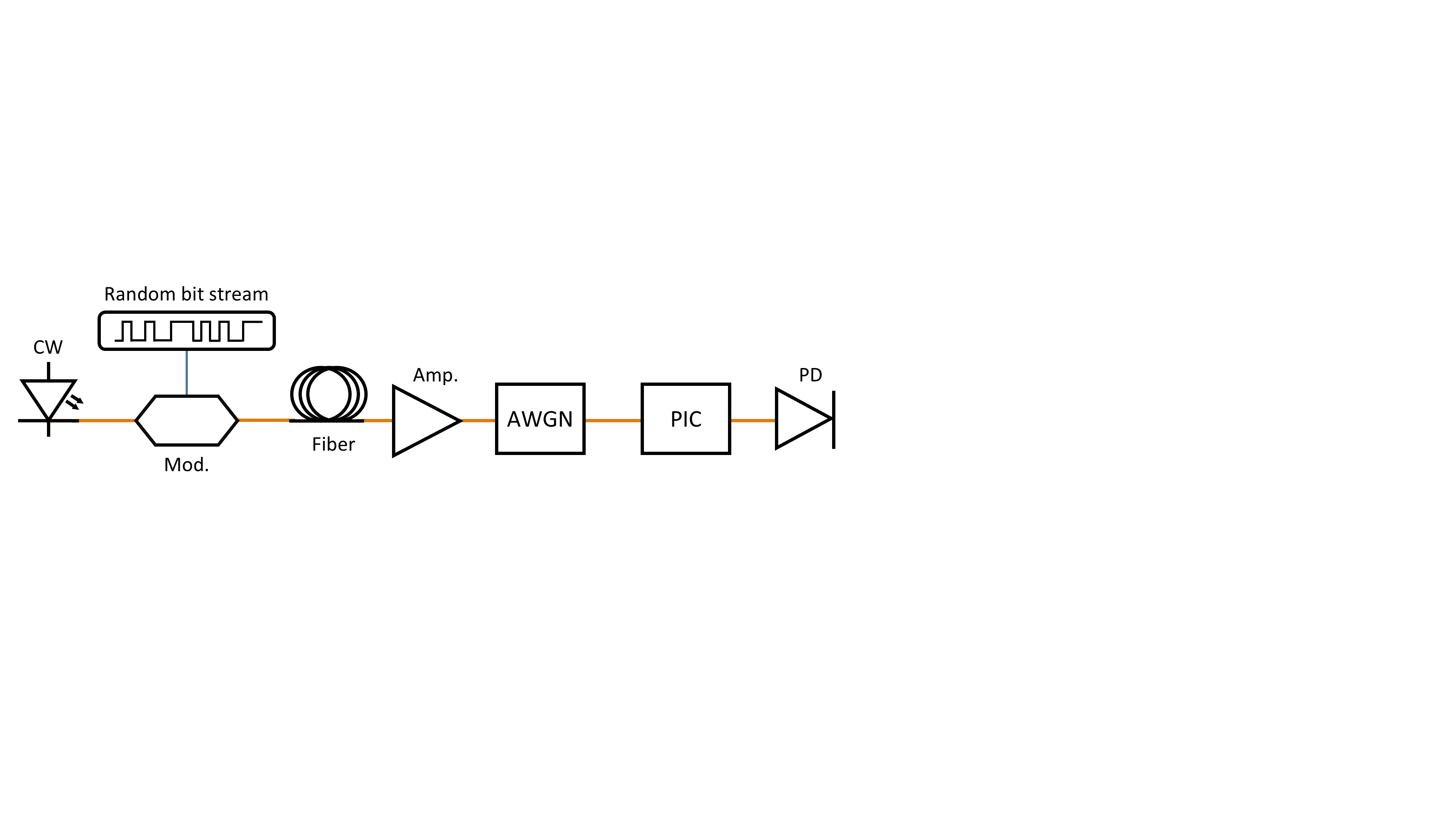}

\caption{Simulation schematic for the transmission. CW: Continuous Wave Laser, Mod.: Modulator, Amp.: Amplifier, AWGN: Additive White Gaussian Noise, PIC: Photonic Integrated Circuit, PD: Photodiode}
\label{fig:vpi_schematic}

\end{figure}

 These architectures were used in a system simulation also implemented using VPIPhotonics Desig Suite, with the schematic shown in Figure \ref{fig:vpi_schematic}. An OOK signal with baudrates between 10 and 100 Gbaud was generated and propagated through varying optical fiber lengths. Noise from various components was lumped and introduced as additive white Gaussian noise (AWGN) prior to the photonic equalizers to achieve an OSNR of 25 dB. Polarization effects were not included in the simulation, and noise contributions from the photodiode were also neglected.

To enable a meaningful evaluation of both performance and memory across different baud rates, the fiber lengths are not chosen uniformly but instead selected to yield the same relative pulse broadening, \(r\), with respect to the bit duration. This allows the distortion to be expressed directly in the effective number of bits that must be retained and equalized by the photonic architecture. The relative broadening is defined as

\[
r = \frac{\Delta T}{T_b},
\]
where \(\Delta T\) is the dispersion-induced pulse broadening and \(T_b = 1/B\) is the bit period for baud rate \(B\). Noting that chromatic-dispersion-induced broadening scales as
\[
\Delta T \approx \frac{D \lambda^2}{c}\, k \,B^2L,
\]
where \(D = 17~\mathrm{ps/(nm\cdot km)}\), \(\lambda = 1550~\mathrm{nm}\), and \(k = 1.3\) accounts for the excess bandwidth of the unshaped pulses used. The required fiber length for a target relative broadening \(r\) is obtained by solving for \(L\) for a given baud rate $B$. 

By fixing  \(r\), each baud rate experiences the same amount of pulse spreading relative to the symbol duration, such that the number of interfering symbols, and therefore the memory requirement of the equalizer, is held constant across baud rates. This enables a direct comparison of how effectively different architectures process a given number of bits of temporal distortion, independent of baud rate. For example, for a target relative broadening of \(r=5\), the corresponding fiber lengths are approximately 282 km at 10 Gbaud, 70.6 km at 20 Gbaud, and 17.6 km at 40 Gbaud, and 2.82 km at 100 Gbaud. 

To maintain consistency with the experiment, readout states are treated as unseen, and the same CMA-ES training procedure is used. Therefore the same processing and optimization steps as in Figure \ref{fig:processing_optimization} are followed. The test results are then evaluated over 128,000 bits and statistical BER estimations from these sets were used. 

\begin{figure}[H]
\centering\includegraphics[width=\textwidth, trim=0cm 4cm 0cm 3cm,clip]{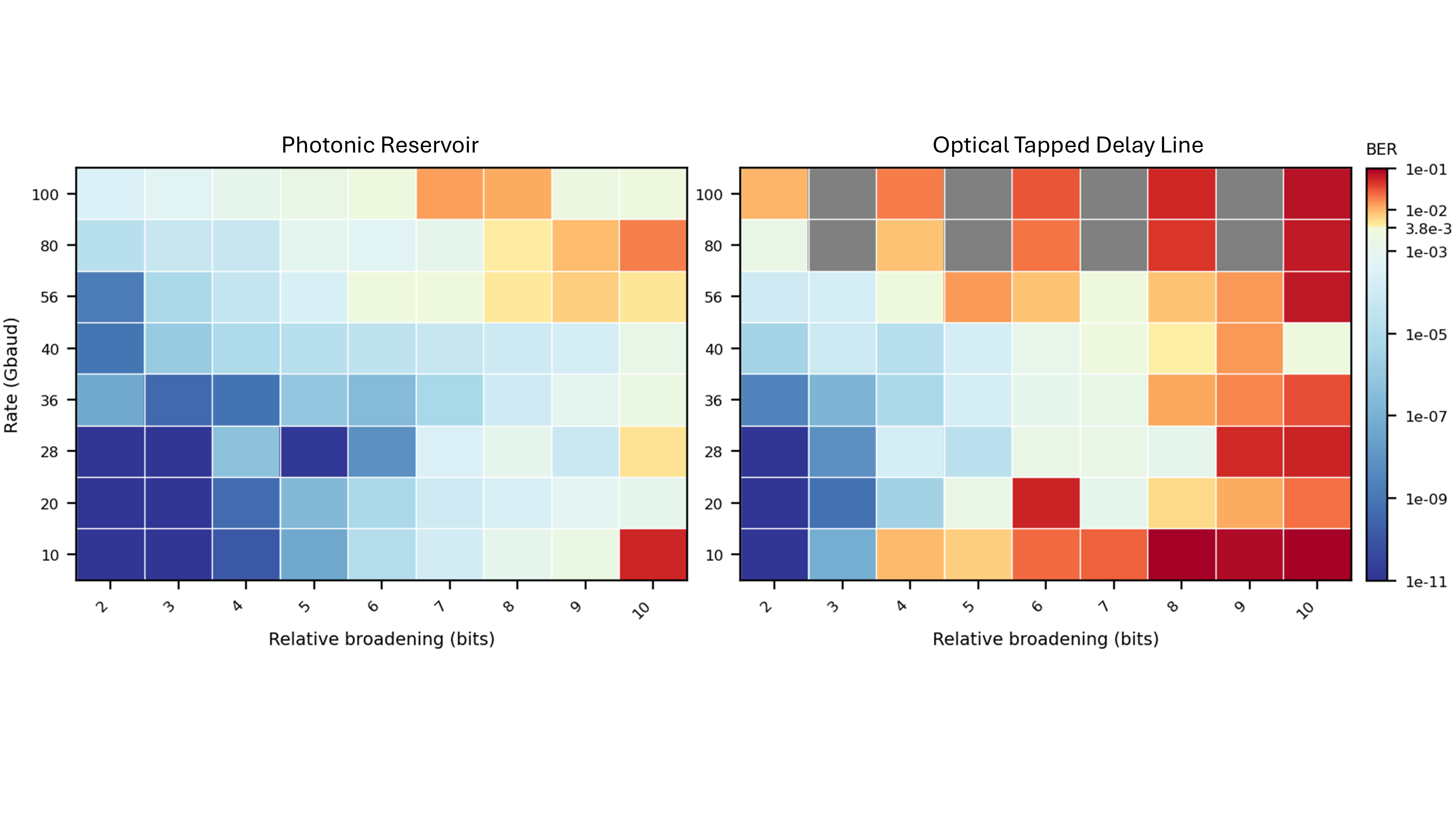}

\caption{Comparison of equalization performance for the photonic reservoir and optical tapped delay line (OTDL) as a function of relative pulse broadening (2–10 bits) across baud rates from 10 to 100 Gbaud. Colors transitioning from blue to green indicate BER below the HD-FEC threshold ($3.8\times10^{-3}$), while colors from orange to red indicate BER above this threshold. Grey cells correspond to configurations that were not simulated due to consistently high BER and limited additional insight.}
\label{fig:main_simulations}

\end{figure}

These results are shown in Fig. \ref{fig:main_simulations} for both the reservoir and the OTDL. Baud rates from 10 to 100 Gbaud are evaluated, with relative pulse broadening ranging from 2 to 10 bits. The reservoir demonstrates strong tolerance to varying baud rates, maintaining BER below the HD-FEC threshold of $3.8\times10^{-3}$ for up to 9 to 10 bits of broadening with baud rates below 56 Gbaud. For baudrates above 56 Gbaud, 6 to 7 bits of broadening are tolerated. 

In contrast, the OTDL exhibits limited robustness to baud rate variations, with only values close to the 40 Gbaud design point (i.e., 28 and 36 Gbaud) remaining below the HD-FEC threshold when up to 7 bits of broadening is incurred. Both higher (e.g., 80 Gbaud) and lower (e.g., 10 Gbaud) rates exceed the threshold already at 4 bits, indicating reduced adaptability. For larger broadening values (8–10 bits), the OTDL fails to support any baud rate, reflecting insufficient memory to compensate for the increased temporal distortion.

These results highlight three key strengths of the reservoir architecture: (i) enhanced memory capacity, enabling compensation of longer effective channel responses; (ii) robustness to operating conditions, allowing a single hardware to function across a broad range of baud rates; and (iii) improved equalization performance, consistently outperforming a conventional optical delay-line architecture, including at its nominal design point.

A final set of simulations was performed to evaluate the impact of the parasitic connection shown in Fig \ref{fig:crosstalk} on the equalization performance in the experiment. Two representative rates were considered, at 20 and 40 Gbaud. For each case, the results are shown both without the crosstalk path included in the readout and with the crosstalk path added.

Because the exact relative power levels of the two paths experienced in the experiment cannot be determined with certainty even when component losses are accounted for in the simulation, the parasitic contribution is evaluated over a representative range. Specifically, the relative power between the main signal and the crosstalk path is considered under three conditions: (i) as determined by component losses alone (0 dB additional offset), and (ii) with the crosstalk path increased by 3 dB and (iii) 5 dB, representing scenarios with higher crosstalk or higher reservoir component losses. The results show that the presence of this parasitic path leads to a degradation in performance of up to several orders of magnitude.

It is worth noting that such a connection would not necessarily be detrimental if it were intentionally incorporated into the architecture with an associated trainable weight. In the present case, however, the connection is uncontrolled and parasitic, and therefore introduces an undesired bypass contribution that degrades the equalization performance, consistent with the limitations observed experimentally.

\begin{figure}[H]
\centering\includegraphics[width=\textwidth, trim=0cm 9.5cm 0cm 1.15cm,clip]{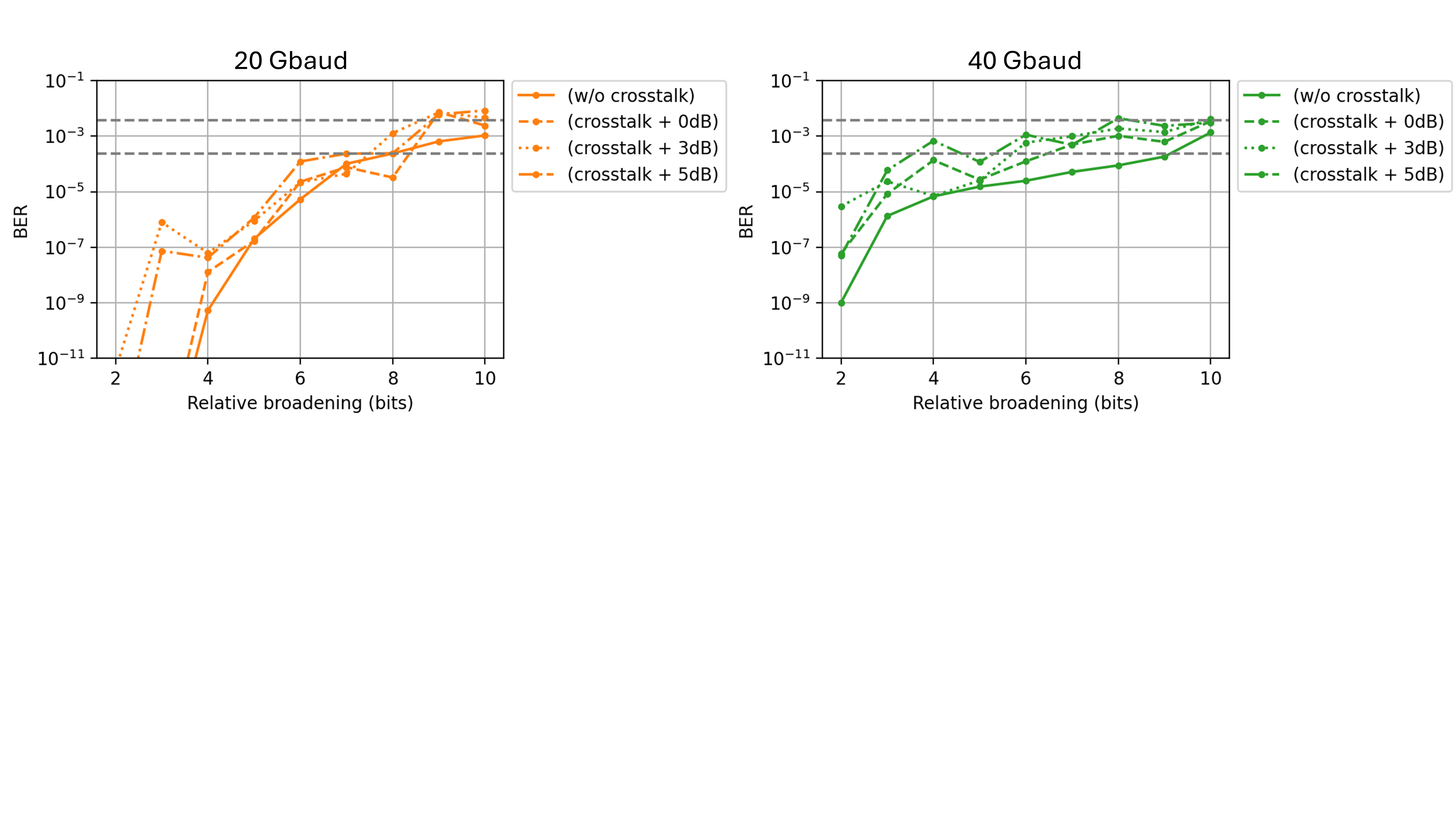}

\caption{Simulated negative impact of the parasitic path on equalization performance at 20 and 40 Gbaud. The relative power of the parasitic contribution is varied with respect to the main signal, expressed as excess over the ratio determined by component losses.}
\label{fig:skip_simulations}

\end{figure}

\subsection{Comparison with the State of the Art}

Recent work has explored photonic machine-learning-based equalization using a range of reservoir computing and neural network architectures. These demonstrations span delay-loop \cite{estebanez56GBaudPAM42022, argyrisPAM4Transmission15502019} and optoelectronic reservoir systems \cite{sozosExperimentalInvestigationRecurrent2024, darosReservoircomputingBasedEqualization2020}, as well as more recent integrated photonic processors with programmable recurrent \cite{asscheRealtimeAllopticalSignal2025,wangTerabitIntegratedNeuromorphic2025} and feed-forward topologies \cite{staffoliRecovery40Gbps2023}. The reported results cover a wide range of transmission scenarios, including OFDM \cite{marcianoChromaticDistortionPrecompensation2025}, coherent \cite{masaadExperimentalDemonstration4Port2024a}, multi-wavelength operation \cite{gooskensExperimentalResultsNonlinear2023, wangTerabitIntegratedNeuromorphic2025}, and high-baud-rate regimes \cite{kuhlCascadedMicroRingResonator2025, wangTerabitIntegratedNeuromorphic2025}.

Despite this progress, several important architectural and experimental gaps remain. First, many demonstrated systems are either not inherently integrable or rely on optoelectronic processing, and therefore do not fully realize all-photonic equalization. Second, for photonic integrated solutions, experimental demonstrations remain limited, and within these, implementations relying on hardware (on-chip) training rather than off-chip optimization are particularly scarce. In this context, the presented demonstrator, which co-integrates the photonic equalizer with the receiver and performs on-chip equalization, represents a significant step toward a practical and experimentally validated solution. To the best of our knowledge, this is the first experimental demonstration of a photonic equalizer co-integrated with the receiver, enabling on-chip equalization in a fully integrated configuration.

Furthermore, the application space of photonic equalizers remains largely divided between single-condition, reach-flexible, or baud-rate-flexible demonstrations. This is illustrated in Table \ref{tab:sota}, which summarizes representative experimental works and categorizes their demonstrated flexibility. While these results represent significant progress and highlight the potential of photonic processing, they do not fully exploit the programmability required for practical and adaptable deployment.To the best of our knowledge, this work is the first to experimentally demonstrate fully photonic equalization with simultaneous flexibility across both transmission reach and baud rate.

\begin{table*}[htbp]
\centering
\caption{Comparison of experimental photonic equalization works.}
\label{tab:sota}
\footnotesize
\renewcommand{\arraystretch}{1.2}

\begin{tabular}{|p{1.2cm}|p{1.2cm}|p{0.9cm}|p{1cm}|p{1.1cm}|p{1.2cm}|p{1.5cm}|p{1.8cm}|}
\hline

\textbf{Year / ref.} & \textbf{Arch.}& \textbf{Format}& \textbf{Rate} (Gbaud)& \textbf{Reach} (km)& \textbf{Integrated }& \textbf{Digital Co-processing}& \textbf{Flexibility} \\
\hline

Argyris 2019 \cite{argyrisPAM4Transmission15502019}&
Delay-loop RC &
PAM-4 &
25, 56 &
21, 4.6 &
No&
Regression with 20 taps per node &
Dual conditions\\
\hline

Est{\'e}banez 2021 \cite{estebanez56GBaudPAM42022}&
Delay-loop RC &
PAM-4 &
56 &
100 &
No&
Regression with 31--61 taps per node &
Single condition \\
\hline

Shen 2023 \cite{shenDeepPhotonicReservoir2023}&
Deep RC &
OOK &
25 &
50 &
No&
Regression &
Single condition \\
\hline

Staffoli 2025 \cite{staffoliNonlinearDistortionEqualization2025}&
PNN (8-node) &
OOK &
10 &
up to 200 &
Yes (SOI, no rx)&
\textbf{None} &
Reach-flexible \\
\hline

Staffoli 2025 \cite{staffoliSiliconPhotonicNeural2025}&
PNN (8-node) &
PAM-4 &
10 &
up to 125 &
Yes (SOI, no rx)&
\textbf{None} &
Reach-flexible \\
\hline

Van Assche 2025 \cite{asscheRealtimeAllopticalSignal2025}&
Spatial RC (8-node) &
OOK &
28 &
up to 50 &
Yes (SiN, no rx)&
\textbf{None} &
Reach-flexible \\
\hline

Teofilovic 2025 \cite{teofilovicIntegratedRecurrentOptical}&
Spectrum-slicing &
PAM-4 &
32 &
up to 100 &
Yes (Si, no rx)&
FFE &
Reach-flexible \\
\hline

Wang 2025 \cite{wangTerabitIntegratedNeuromorphic2025}&
Deep delay RC&
OOK, PAM-4 &
\textbf{56 -- 112} &
5 &
Yes (SOI, no rx)&
\textbf{None} &
Baud-flexible \\
\hline

Masaad 2026 (this work)&
Spatial RC (16-node)&
OOK &
\textbf{10 -- 46} &
\textbf{up to 250} &
\textbf{Yes (SiN, rx on InP)}&
\textbf{None} &
\textbf{Baud + reach flexible }\\
\hline

\end{tabular}
\begin{minipage}{\textwidth}
\scriptsize
\textit{Abbreviations:} Arch.: architecture; RC: reservoir computing; 
PNN: photonic neural network; OOK: on--off keying; PAM-4: four-level 
pulse amplitude modulation; FFE: feed-forward equalizer.
\end{minipage}
\end{table*}

\section{Conclusion}
We have experimentally demonstrated a co-packaged, receiver-side photonic reservoir computing equalizer capable of operating across a wide range of transmission conditions without modification of the underlying photonic circuit. Using a fixed-topology integrated reservoir with programmable optical readout, equalization was achieved across baudrates from 10 to 46 Gbaud and standard single-mode fiber reaches up to 250 km in the C-band. Adaptation across these regimes was obtained solely through retraining of the readout weights, leveraging the recurrent dynamics and fading memory of the reservoir rather than hardware reconfiguration.

Across the evaluated scenarios, the reservoir equalizer consistently provided multi-order-of-magnitude BER improvement relative to a linear DSP feed-forward equalizer baseline. These results establish that integrated photonic reservoirs can function as flexible, reusable equalization processors beyond single-operating-point devices typical in the field. The demonstrated baudrate and reach generalization with fixed hardware supports a deployment model in which a single co-packaged photonic processor can be retrained for multiple link configurations. This positions photonic reservoir equalization as a practical replacement to DSP for receiver-side signal conditioning in power- and latency-constrained optical links.

\begin{backmatter}
\bmsection{Funding}
Parts of this work were performed in the context of the European projects Nebula (GA871658), Prometheus (GA101070195), Nehil (GA101194363), Neho (GA101046329), and Neuropuls (GA101070238). The work was also supported by an FWO travel grant (V431325N)

\bmsection{Acknowledgment}

\bmsection{Disclosures}
The authors declare no conflicts of interest.

\bmsection{Data Availability Statement}

\end{backmatter}
\bibliography{test}
\end{document}